\documentclass[preprint,showpacs,preprintnumbers,amsmath,amssymb]{revtex4}

\usepackage{graphicx}
\usepackage{dcolumn}
\usepackage{bm}

\begin{document}

\preprint{}

\title{Topological structure of the vortex solution in Jackiw-Pi model}
\author{LEE Xi-Guo$^{1,3}$}
\email{xgl@impcas.ac.cn;xiguoli@ns.lzb.ac.cn}
\author{LIU Zi-Yu$^{1,2}$}
\author{LI Yong-Qing$^{1,2}$}
\author{GAO Yuan$^{1,2}$}
\author{GUO Yan-Rui $^{1,2}$}
\author{XIAO Guo-qing $^{1}$}

\affiliation{$^{1}$Institute of Modern Physics, Chinese Academy of
Science, P.O. Box 31, Lanzhou 730000, People's republic of china}
\affiliation{$^{2}$Graduate School, Chinese Academy of Science,
Beijing 100049, People's republic of china}
\affiliation{$^{3}$Center of Theoretical Nuclear Physics, National
Laboratory of Heavy Ion Collisions, Lanzhou 730000, China}

\begin{abstract}
By using $\phi$ -mapping method, we discuss the topological
structure of the self-duality solution in Jackiw-Pi model in terms
of gauge potential decomposition. We set up relationship between
Chern-Simons vortex solution and topological number which is
determined by Hopf index and Brouwer degree. We also give the
quantization of flux in this case. Then, we study the angular
momentum of the vortex, it can be expressed in terms of the flux.

\end{abstract}

\maketitle

\section{introduction}
Chern-Simons theories exhibit many interesting and important
properties. They are based on secondary characteristic classes
discovered in Ref.\cite{cs01} and many topological invariants of
knots and links discovered in the 1980s could be reinterpreted as
correlation functions of Wilson loop operators in Chern-Simons
theory\cite{wt03}. Moreover, for gauge theories and gravity in
three-dimensions, they can appears as natural mass terms and will
lead to a quantized coupling constant as well as a mass after
quantization\cite{dj02}. They have also found applications to a
lot of physical problems, such as particle physics, quantum Hall
effect, quantum gravity and string
theory\cite{ykf,ba98,jp01,re01,qh01,re02,re03,re04,re05,wt01,wt02}.
 Chern-Simons term acquire dynamics via
coupling to other fields\cite{ah01,ba98}, and get multifarious
gauge theory, all of which have vortex solutions, these static
solutions can
 be obtained when their Hamiltonian was minimal. Vortices and their dynamics are interesting objects
 to be studied\cite{ah01,pd02,pd23,pd33,pd43}. R. Jackiw and S-Y. Pi
considered a gauged, nonliner Sch\"{o}dinger equation in two
spatial dimensions, with describes nonrelativistic  matter
interacting with Chern-Simons gauge fields. Then they find
explicit static, self-dual solutions which satisfies the Liouville
equation.

In this paper, we will discuss topological structure of the
self-duality solution in Jackiw-Pi model in terms of gauge
potential decomposition\cite{dg01,du01,sa01,li04,li05}. We will
look for complete vortex solution from self-duality equation and
set up the relationship between the vortex solution and
topological number. We also study the quantization of the flux of
the vortex. Last, we will investigate the angular momentum of the
vortex.

\section{Self-duality solutions in Jackiw-Pi Model}
In this section, making use of the self-duality equation, we will
look for complete vortex solution in Jackiw-Pi model\cite{ba98} by
using the decomposition of gauge potential. The Abelian Jackiw-Pi
model in nonlinear Schr\"odinger systems is
\begin{eqnarray}
\pounds_{jp}=\frac{\kappa}{2}\epsilon^{\mu\nu\lambda}A_{\mu}\partial_{\nu}A_{\lambda}+i\hbar\Psi^{*}D_{0}\Psi-\frac{\hbar^{2}}{2m}|\mathbf{D}\Psi|^{2}+\frac{g}{2}(\Psi^{*}\Psi)^{2}.
\end{eqnarray}
[We use relativistic notation with the metric diag(1,-1,-1) and
$x^{\mu}=(ct,\mathbf{r})$.] Where
$\mathbf{D}=\mathbf{\nabla}-i\frac{e}{\hbar c}\mathbf{A}$ and
$\Psi$ is "matter" field, the first term is the Chern-Simons
density, which is not gauge invariant. Also $m$ is the mass
parameter, $A_{\mu}$ is gauge potentials, $g$ governs the strength
of nonlinearity, $\kappa$ controls the Chern-Simons addition and
provides a cutoff at large distance greater than
$\frac{1}{\kappa}$ for gauge-invariant electric and magnetic
fields, which can be written as $\mathbf{E}=-\nabla
\mathbf{A}^{0}-(\frac{1}{c})\partial_{t}\mathbf{A}$, and
$\mathbf{B}=\mathbf{\nabla}\times\mathbf{A}$. Thus the
Chern-Simons terms gives rise to massive, yet gauge-invariant
"electrodynamics". The last term represents a self-coupling
contact term of the type commonly found in nonlinear Schr¡§odinger
systems. The Euler-Lagrange equations are
\begin{eqnarray}
iD_0
\Psi=-\frac{1}{2m}\mathbf{D}^{2}\Psi-g\left|\Psi\right|^2\Psi,
F_{\mu\nu}=\frac{1 }{\kappa} \varepsilon_{\mu\nu\rho} J^{\rho}.
\end{eqnarray}
The energy density is
\begin{eqnarray}
\mathcal{H}=\frac{\hbar^{2}}{2m}|\mathbf{D}\Psi|^{2}-\frac{g}{2}(\Psi^{*}\Psi)^{2},
\end{eqnarray}
in which $g=\mp\frac{e^{2}\hbar}{mck}$, by
\begin{eqnarray}
|\mathbf{D}\Psi|^{2}=|(D_{1}\pm
iD_{2})\Psi|^{2}\pm\frac{m}{\hbar}\mathbf{\nabla}\times\mathbf{J}\pm\frac{e}{\hbar
c}B\Psi^{*}\Psi,
\end{eqnarray}
and
\begin{eqnarray}
B=\epsilon^{ij}\partial_{i}A^{j}=-\frac{e}{k}\rho,
\end{eqnarray}
we can get
\begin{eqnarray}
\mathcal{H}=\frac{\hbar^{2}}{2m}|(D_{1}\pm
iD_{2})\Psi|^{2}\pm\frac{\hbar}{2}\nabla\times\mathbf{J}-\left[\frac{g}{2}\pm\frac{e^{2}\hbar}{2mck}\right](\Psi^{*}\Psi)^{2}.
\end{eqnarray}
where $e$ measure the coupling to gauge field described by scalar
$A^{0}$ and vector $\mathbf{A}$ potentials. With
$g=\mp\frac{e^{2}\hbar}{mck}$, and sufficiently well-behaved
fields so that the integral over all space of
$\nabla\times\mathbf{J}$ vanishes, the energy is
\begin{eqnarray}
H=\int d\mathbf{r}\mathcal{H}=\frac{\hbar^{2}}{2m}\int
d\mathbf{r}|(D_{1}\pm iD_{2})\Psi|^{2}.
\end{eqnarray}
This is non-negative and vanishes. Thus attaining its minimum when
$\Psi$ satisfies a self-dual equation
\begin{eqnarray}
D_{1}\Psi=\mp iD_{2}\Psi.
\end{eqnarray}
To solve Eq.(8),we note that when $\Psi$ is decomposed into two
scalar fields
\begin{eqnarray}
\Psi=\Psi_{1}+i\Psi_{2},
\end{eqnarray}
We can define a unit vector field $\mathbf{n}$ as follows
\begin{eqnarray}
n^{a}=\frac{\Psi_{a}}{(\Psi^{*}\Psi)^{\frac{1}{2}}},a=1,2.
\end{eqnarray}
It is easy to prove that $\mathbf{n}$ satisfies the constraint
conditions
\begin{eqnarray}
n^{a}n^{a}=1,    n^{a}dn^{a}=0.
\end{eqnarray}
From Eq.(8), and making use of the decomposition of U(1) gauge
potential in terms of the two-dimensional unit vector field
$\mathbf{n}$, we can obtain
\begin{eqnarray}
A^{i}=\frac{\hbar
c}{e}(\epsilon^{ab}n^{a}\partial_{i}n^{b}\pm\frac{1}{2}\epsilon^{ij}\partial_{j}\ln\rho),
\end{eqnarray}
From Eq.(5) and Eq.(12) we get
\begin{eqnarray}
B=\frac{\hbar
c}{e}\epsilon^{ij}\epsilon^{ab}\partial_{i}n^{a}\partial_{j}n^{b}\pm\frac{\hbar
c}{2e}\nabla^{2}\ln\rho,
\end{eqnarray}
i.e.
\begin{eqnarray}
-\frac{e}{k}\rho=\frac{\hbar
c}{e}\epsilon^{ij}\epsilon^{ab}\partial_{i}n^{a}\partial_{j}n^{b}\pm\frac{\hbar
c}{2e}\nabla^{2}\ln\rho.
\end{eqnarray}
This equation can be rewritten as
\begin{eqnarray}
\nabla^{2}\ln\rho=\pm\frac{2e^{2}}{\hbar c
k}\rho\pm2\epsilon^{ij}\epsilon^{ab}\partial_{i}n^{a}\partial_{j}n^{b}.
\end{eqnarray}
With the help of the $\phi$ -mapping method\cite{du01,sa01},
Eq.(15) can be written as
\begin{eqnarray}
\nabla^{2}\ln\rho=\pm\frac{2e^{2}}{\hbar c
k}\rho\pm4\pi\delta^{2}(\mathbf{\Psi})J\left(\frac{\mathbf{\Psi}}{\mathbf{x}}\right),
\end{eqnarray}
in which $J\left(\frac{\mathbf{\Psi}}{\mathbf{x}}\right)$ is
Jacobian and
\begin{eqnarray}
J\left(\frac{\mathbf{\Psi}}{\mathbf{x}}\right)=\frac{1}{2}\epsilon^{ab}\epsilon^{ij}\frac{\partial\Psi_{a}}{\partial
x^{i}}\frac{\partial\Psi_{b}}{\partial x^{j}},       i,j=1,2.
\end{eqnarray}
When $\rho\neq0$, Eq.(16) will be the Liouville equation,
\begin{eqnarray}
\nabla^{2}\ln\rho=\pm\frac{2e^{2}}{\hbar c k}\rho,
\end{eqnarray}
as we all know, the Eq.(18) has the general real solution as
follows
\begin{equation}
\rho=\kappa \nabla^2\ln\left(1+|f|^2\right),
\end{equation}
where $f=f(z)$ is a holomorphic function of $z=x^1+ix^2$ only.

Using $f(z)=(\frac{z_0}{z})^{N}$, it is easy to obtain the
radially symmetric solutions with $r\neq0$ \cite{ax01}
\begin{equation}
\rho=\pm\frac{4 N^2}{ r_0^2}\frac{\hbar
c\kappa}{e^{2}}\frac{\left(\frac{r}{r_0}\right)^{2(N-1)}}{\left(1+\left(\frac{r}{r_0}\right)^{2N}\right)^2}
.
\end{equation}
Because $\rho$ is the charge density of the vortex, it must be
positive, the Liouville equation is
\begin{eqnarray}
\nabla^{2}\ln\rho=-\frac{2e^{2}}{\hbar c|\kappa|}\rho,
\end{eqnarray}
so the Eq.(20) should be
\begin{eqnarray}
\rho=\frac{\hbar c}{e^{2}}\frac{4|\kappa|
N^{2}}{r_{0}^{2}}\frac{\left(\frac{r}{r_{0}}\right)^{2(N-1)}}{
\left(1+\left(\frac{r}{r_{0}}\right)^{2N}\right)^{2}}
\label{radialsol},
\end{eqnarray}
and the Eq.(16) can be rewritten as
\begin{eqnarray}
\nabla^{2}\ln\rho=-\frac{2e^{2}}{\hbar c
|\kappa|}\rho-\frac{\kappa}{|\kappa|}4\pi\delta^{2}(\Psi)J\left(\frac{\mathbf{\Psi}}{\mathbf{x}}\right).
\end{eqnarray}
\section{the topological structure of the vortex solution and its magnetic flux }
In this section, making use of Eq.(23), we will discuss the
topological structure of the vortex solution, then we will study
the magnetic flux of the vortex. Under the radially symmetric,
$\nabla^{2}\ln\rho$ can be expressed as
\begin{eqnarray}
\nabla^{2}\ln\rho=\frac{\partial^{2}}{\partial
^{2}r}\ln\rho+\frac{1}{r}\partial_{r}\ln\rho.
\end{eqnarray}
Substituting Eq.(22) into Eq.(24), we can get
\begin{eqnarray}
\nabla^{2}\ln\rho=-\frac{8N^{2}(\frac{r}{r_{0}})^{2N-2}}{r_{0}^{2}[1+(\frac{r}{r_{0}})^{2N}]^{2}}+2(N-1)\nabla\left(\frac{\mathbf{r}}{r^{2}}\right).
\end{eqnarray}
Integrating Eq.(23)
\begin{eqnarray}
\int\nabla^{2}\ln\rho d\mathbf{r}=\int\left[-\frac{2e^{2}}{\hbar c
|\kappa|}\rho-\frac{\kappa}{|\kappa|}4\pi\delta^{2}(\mathbf{\Psi})J\left(\frac{\mathbf{\Psi}}{\mathbf{x}}\right)\right]d\mathbf{r}.
\end{eqnarray}
Suppose that the vector field $\Psi^{a}$ possess one isolated
zeros which is in $x=0$, according to the $\delta$-function
theory\cite{delt}, $\delta^{2}(\mathbf{\Psi})$ can be expressed by
\begin{equation}
\delta^{2}(\mathbf{\Psi})=\frac{\beta \delta^{2}(\mathbf{x})}{\mid
J\left(\frac{\mathbf{\Psi}}
{\mathbf{x}}\right)\mid_{\mathbf{x}=0}},
\end{equation}
and then one can obtain
\begin{eqnarray}
\int4\pi\delta^{2}(\mathbf{\Psi})J\left(\frac{\mathbf{\Psi}}{\mathbf{x}}
\right)d\mathbf{r}=4\pi \int
\beta\eta\delta^{2}(\mathbf{x})d\mathbf{r},
\end{eqnarray}
where $\beta$ is positive integer (the Hopf index of the zero
point) and $\eta$, the Brouwer degree of the vector field
$\mathbf{\Psi}$,
\begin{eqnarray}
\eta=sgnJ\left(\frac{\mathbf{\Psi}}{\mathbf{x}}\right)|_{\mathbf{x=0}}=\pm1.
\end{eqnarray}
The meaning of the Hopf index $\beta$ is that while $\mathbf{x}$
covers the region neighbouring the zero point once, the vector
field $\mathbf{\Psi}$ covers the corresponding region $\beta$
times, Hence, $\beta$ and $\eta$ are the topological number which
shows the topological properties of the vortex solution. We have
\begin{eqnarray}
\delta^{2}(\mathbf{\Psi})J\left(\frac{\mathbf{\Psi}}{\mathbf{x}}\right)=\beta\eta\delta^{2}(\mathbf{x}).
\end{eqnarray}
If we define the topological number $Q$ as
\begin{eqnarray}
Q=\int\delta^{2}(\mathbf{\Psi})J\left(\frac{\mathbf{\Psi}}{\mathbf{x}}\right)d\mathbf{r}=\beta\eta,
\end{eqnarray}
from Eq.(26) we can get
\begin{eqnarray}
N-1=-\frac{\kappa}{|\kappa|}Q.
\end{eqnarray}
Substituting Eq.(32) into Eq.(22), we can obtain
\begin{eqnarray}
\rho=\frac{\hbar
c}{e^{2}}{4|\kappa|(-\frac{\kappa}{|\kappa|}Q+1)^2\over
r_0^2}{\left(\frac{r}{r_0}\right)^{-2\frac{\kappa}{|\kappa|}Q}\over
\left(1+\left(\frac{r}{r_0}\right)^{2(-\frac{\kappa}{|\kappa|}Q+1)}\right)^2},
\end{eqnarray}
It is obviously Eq.(33) is the solution of Eq(23). On the other
hand, this means vortex density $\rho$ relates to its topological
number $Q$. We now see that $N$ must be an integer.

If we note the unit magnetic flux $\Phi_{0}=\frac{2\pi\hbar
c}{e}$, one can get
\begin{eqnarray}
\int_{0}^{\infty}
\mathbf{B}d\mathbf{r}=-4\pi\kappa\left|\frac{N}{\kappa}\right|\frac{\hbar
c}{e}=-2\frac{\kappa}{|\kappa|}\Phi_{0}\left|-\frac{\kappa}{|\kappa|}Q+1\right|,
\end{eqnarray}
from this equation we know the magnetic flux is quantized. When
the total topological charge equal to zero, the magnetic flux of
this vortex is
\begin{eqnarray}
\Phi=\int_{0}^{\infty}
\mathbf{B}d\mathbf{r}=-2\frac{\kappa}{|\kappa|}\Phi_{0}.
\end{eqnarray}
See Figure [1] for a plot of the density with the $Q=0$ case, and
Figure [2] with the $Q=1$ case. Note the ring-like form of the
magnetic field for these Chern-Simons vortices, as the magnetic
field is proportional to $\rho$, so $\mathbf{B}$ vanishes where
the field $\Psi$ vanishes.
 We define $r_{v}$ is the radius of the vortex, which satisfies
\begin{eqnarray}
\rho_{max}=\rho(r_{v}),
\end{eqnarray}
so $r_{v}$ is the solution of the equation
\begin{eqnarray}
\frac{\partial\rho}{\partial r}=0,
\end{eqnarray}
hence
\begin{eqnarray}
r_{v}=r_{0}\left(1-\frac{2}{-\frac{\kappa}{|\kappa|}Q+1}\right)^{\frac{1}{-2\frac{\kappa}{|\kappa|}Q+2}}.
\end{eqnarray}
In Figure [1] and Figure [2], the radius of the vortex is
$\frac{r_{v}}{r_{0}}$,  and the height of the vortex is
 $\rho_{max}$.  Figure [3] shows the values of the radius of the vortex as $Q$ is
varied when $\kappa <0$.

The height of the vortex
\begin{eqnarray}
\rho_{max}=\frac{\hbar
c|\kappa|}{r_{0}^{2}e^{2}}\left(\frac{\frac{\kappa}{|\kappa|}Q}{\frac{\kappa}{|\kappa|}Q-2}\right)^{\frac{1}{\frac{\kappa}{|\kappa|}Q-1}}\left(\left(\frac{\kappa}{|\kappa|}Q-1\right)^{2}-1\right).
\end{eqnarray}
Figure [4] shows the values of the height of the vortex as $Q$ is
varied when $\kappa <0$.
\section{the angular momentum  of the vortex}
The density $\mathcal{J}$ for the angular momentum $\mathbf{J}$
is\cite{jp01}
\begin{eqnarray}
\mathcal{J}=m\mathbf{r}\times\mathbf{j},
\end{eqnarray}
in which
\begin{eqnarray}
\mathbf{j}=\mp\frac{\hbar}{2m}\nabla\times\rho,
\end{eqnarray}
so we can obtain the angular momentum
\begin{eqnarray}
\mathbf{J}=m\int
d\mathbf{r}\left[\mathbf{r}\times\left(\mp\frac{\hbar}{2m}\nabla\times\rho\right)\right]=\pm\frac{2\hbar
k}{e}\left(\frac{-\kappa}{|\kappa|}Q+1\right)\Phi_{0}=\pm\frac{\kappa\hbar}{e}\Phi.
\end{eqnarray}
where topological number $Q$ must be an integer. This equation
shows $J$ is the magnetic dipole moment.
\section{Conclusions}
When added the usual Maxwell action to Chern-Simons, the resulting
theory represents a single local degree of freedom, paradoxically
endowed with a finite range but still gauge invariant. In this
paper, we studied the topological structure of Chern-Simons vortex
in Jakiw-Pi model. We also obtain the charge of the vortex which
is determined by Hopf index and Brouwer degree. Secondly, compare
with Jakiw's results, we get a Liouville equation with a $\delta$
function, then we also obtain the solution of this equation, and
the $\delta$ function will not change the character of the
solution when $\rho\neq0$. Lastly, we calculate the integral of
the Liouville equation and find the relationship between
topological number and the solution of Liouville equation,
 we also find that the flux is quantized from the integral value of the
solution in the whole space. However, the flux is non-vanish when
the topological number equals to zero. So does the the angular
momentum.

\section{acknowledgments}
This work was supported by the CAS Knowledge Innovation Project
(No.kjcx2-sw-No2;No.kjcx2-sw-No16) and Science Foundation of China
(10435080, 10275123).

\section{references}

 \newpage
\begin{figure}[ht]
  \begin{center}
    \rotatebox{0}{\includegraphics*[width=0.7\textwidth]{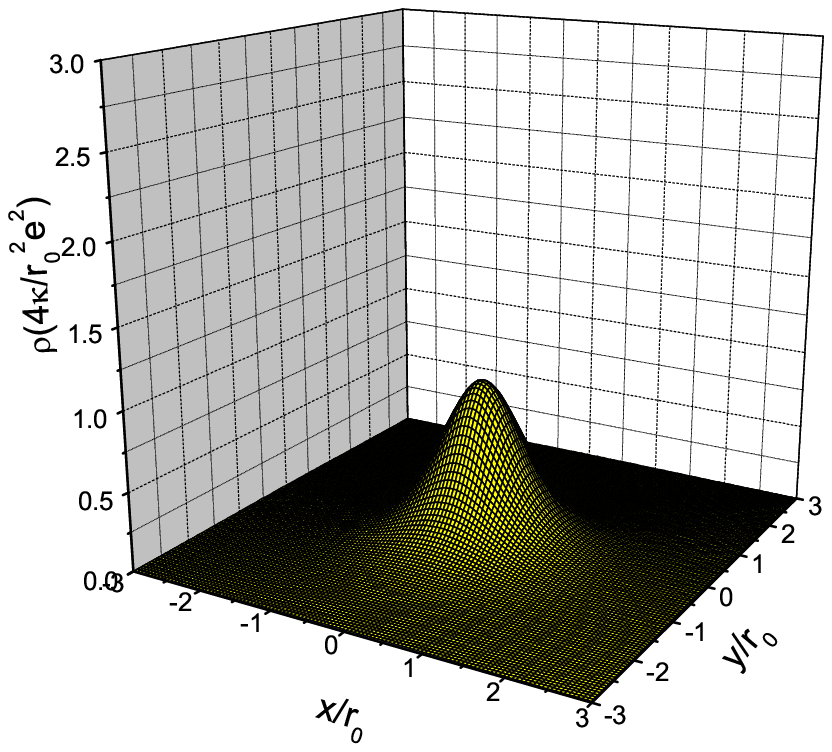}}
    \caption{Density $\rho$ for a radially symmetric solution (33) representing one vortex with $Q=0$.($x\neq0, y\neq0, \hbar=c=1, \kappa>0$).}
  \end{center}
\end{figure}

\begin{figure}[ht]
  \begin{center}
    \rotatebox{0}{\includegraphics*[width=0.7\textwidth]{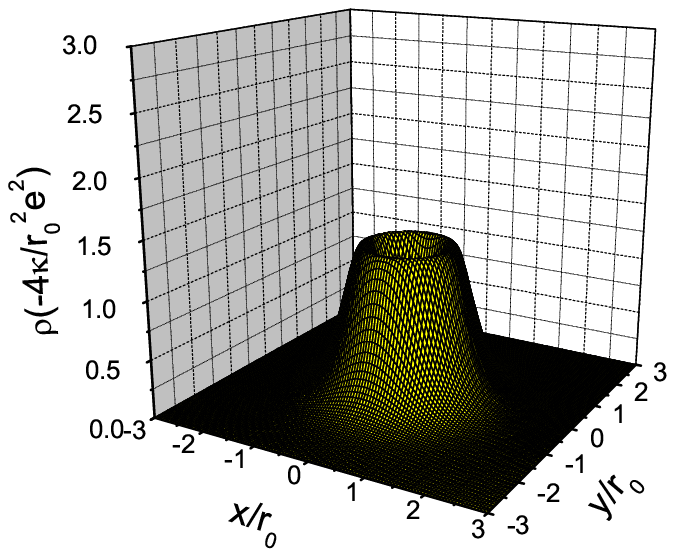}}
    \caption{Density $\rho$ for a radially symmetric solution (33) representing one vortex with $Q=1$.($\kappa<0, \hbar=c=1$)}
  \end{center}
\end{figure}

\newpage
\begin{figure}[ht]
  \begin{center}
    \rotatebox{0}{\includegraphics*[width=0.7\textwidth]{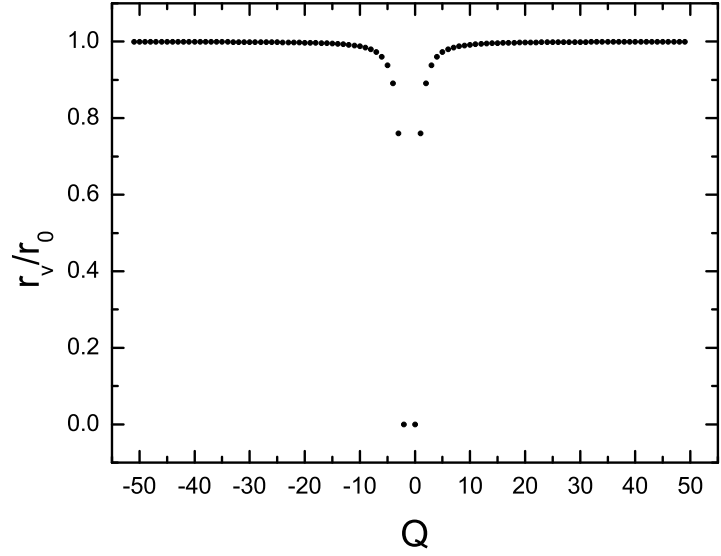}}
    \caption{The values of radius of the vortex as $Q$ is varied when $\kappa
<0$.($\hbar=c=1$)}
  \end{center}
\end{figure}

\begin{figure}[ht]
  \begin{center}
    \rotatebox{0}{\includegraphics*[width=0.7\textwidth]{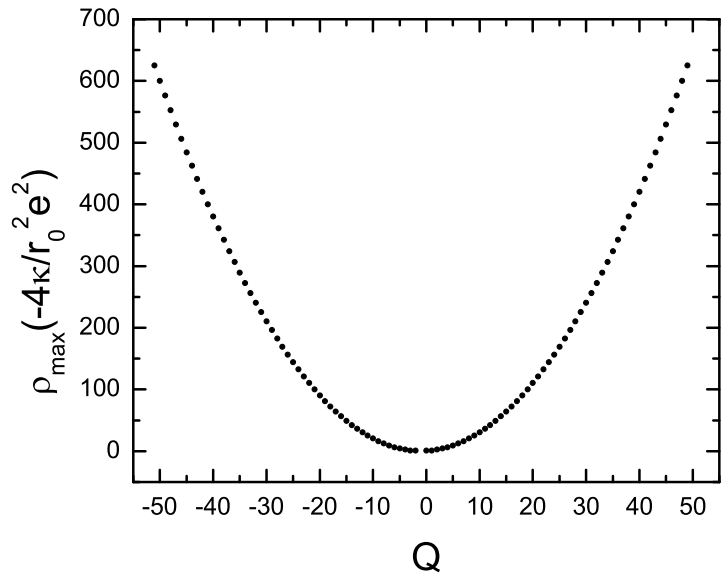}}
    \caption{The values of the height of the vortex as $Q$ is varied when $\kappa
<0$.($\hbar=c=1$)}
  \end{center}
\end{figure}

\end{document}